 \documentclass[preprint2]{aastex}

\usepackage{natbib}
\usepackage{graphicx}
\slugcomment{}

\shorttitle{The PLATO Simulator: Realistic Modelling of Space-Based Imaging}
\shortauthors{Marcos-Arenal et al.}

\begin{document}

%% one line. However, you may use  to force a line break if

\title{The ASTROID Simulator Software Package: Realistic Modelling of High-Precision High-Cadence Space-Based Imaging}

%% As in the title, use  to force line breaks.

\author{P. Marcos-Arenal\altaffilmark{1}, W. Zima\altaffilmark{1},  J. De Ridder\altaffilmark{1}, R. Huygen\altaffilmark{1}, C. Aerts\altaffilmark{1, 2}}
\affil{\altaffilmark{1}Instituut voor Sterrenkunde, KU Leuven, Celestijnenlaan 200D, 3001 Leuven, Belgium}
\affil{\altaffilmark{2}Department of Astrophysics, IMAPP, Radboud University Nijmegen, 6500 GL Nijmegen, The Netherlands}
\email{pablo.MarcosArenal@ster.kuleuven.be}


%\altaffiltext{1}{Visiting Astronomer, Cerro Tololo Inter-American Observatory.
%CTIO is operated by AURA, Inc.\ under contract to the National Science


\begin{abstract}
The preparation of a space-mission that carries out any kind of imaging to detect high-precision low-amplitude variability of its targets requires a robust model for the expected performance of its instruments. This model cannot be derived from simple addition of noise properties due to the complex interaction between the various noise sources. While it is not feasible to build and test a prototype of the imaging device on-ground, realistic numerical simulations in the form of an end-to-end simulator can be used to model the noise propagation in the observations. These simulations not only allow studying the performance of the instrument, its noise source response and its data quality, but also the instrument design verification for different types of configurations, the observing strategy and the scientific feasibility of an observing proposal. In this way, a complete description and assessment of the objectives to expect from the mission can be derived.
We present a high-precision simulation software package, designed to simulate photometric time-series of CCD images by including realistic models of the CCD and its electronics, the telescope optics, the stellar field, the jitter movements of the spacecraft, and all important natural noise sources. This formalism has been implemented in a software tool, dubbed ASTROID Simulator.

%In order to accomplish the multi-mission task, the simulator has been constructed based on two main ideas: employing an architecture based on modularity principles and mimicking a common science imaging pipeline. The modularity principles permit to treat any of the steps in the processing as kind-of plug-in and add or modify the implemented functionalities. Following a regular common pipeline allows any inexperienced user to understand and modify whatever step in the process is different or has any different feature than its own mission from a standard regular processing.

\end{abstract}



%\tableofcontents

\section{Introduction}

The ASTROID Simulator is a software package for realistic modeling of high-precision high-cadence space-based imaging observations of a selected stellar field. It permits to estimate the impact of instrumental and natural noise sources to predict the quality of data and the performance of the instrument. 
	This tool has been designed to perform simulations for a large range of different set-ups thanks to its flexibility and an extended set of input parameters; the ASTROID Simulator is built to be applicable to space missions with a wide range of detector characteristics. The simulator is currently considering charge coupled device (hereafter CCD) simulations but has been redeveloped in a modular architecture that easily allows modifications and implementation of new functionalities, so it can be used for different kinds of future missions with straightforward modifications.

In order to accomplish the multi-mission task, the simulator has been constructed based on two main ideas: employing an architecture based on modularity principles and mimicking a standard science imaging pipeline. The modularity permits to treat any of the steps in the processing independently, and add or modify the implemented functionalities. Any inexperienced user can easily improve the comprehension of the process due to the standard pipeline architecture. The standard pipeline architecture also makes the access to the source code lighter. For users who want to adapt this simulator to particular space missions, it is easy to identify whatever step in the process is different or has any different feature than those of the present standard regular processing.

%%% Figure 1 %%%%%%%%%%%%%%%%%%%%%%%%%%%%%%%%%%%%%%%%%%%%%%%%%%%%%%%%%%%%%%%%%%%%%%%%%
\begin{figure*}[!ht]
\centering
\includegraphics[width=\textwidth, clip,angle=0]{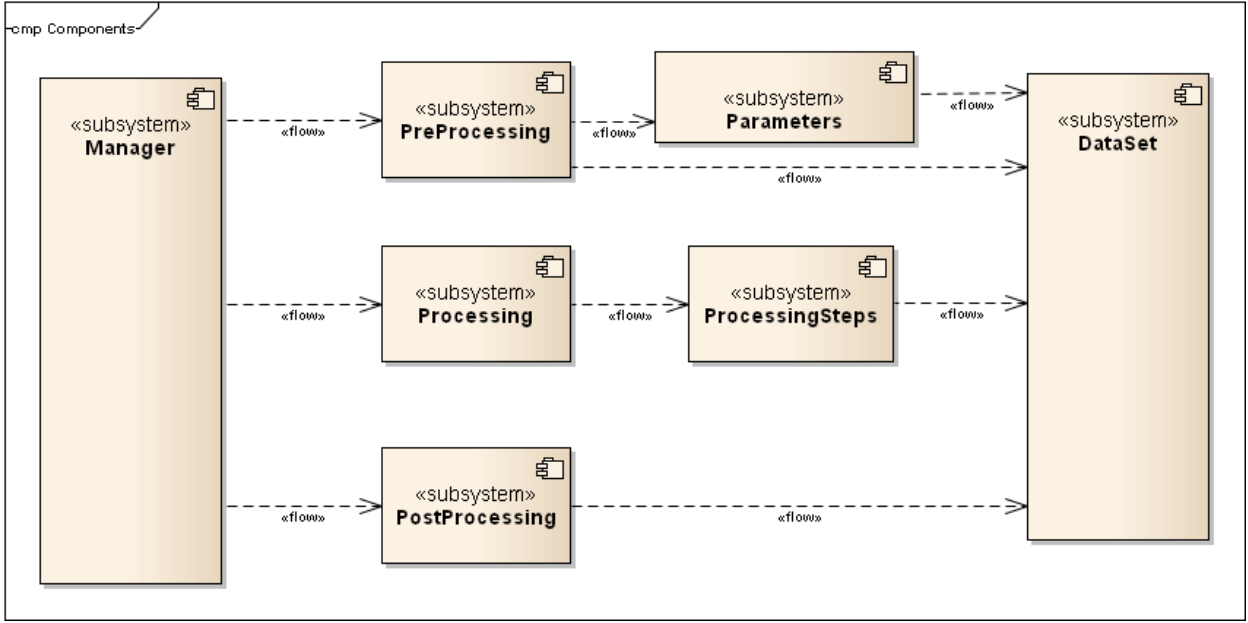}
\caption{Architectural design components relationship overview.}
\label{fig:architecture}
\end{figure*}
%%% Figure 1 %%%%%%%%%%%%%%%%%%%%%%%%%%%%%%%%%%%%%%%%%%%%%%%%%%%%%%%%%%%%%%%%%%%%%%%%%

	A photometry module has been implemented to analyze the simulations. This module analyzes every image generated in the CCD processing module to obtain photometric fluxes and compares the input star catalogue sources with the sources obtained in the produced images. This is described in detail in Sect. 4.
	The ASTROID Simulator is based on the original PLATO Simulator \citep{Zima2010} which was developed for the assessment study of the PLATO M3 mission candidate within ESA's Cosmic Vision 2015-2025 programme. The ASTROID Simulator has been redesigned and implemented as a multi-mission imaging simulator aiming the ease of including new functionalities and detectors. \cite{Catala2008}, \cite{Claudi2010}, \cite{Rauer2010} provide detailed descriptions of the PLATO mission. The PLATO Simulator, in turn, is based on pre-existing codes that were developed for the ESA Eddington mission candidate (\citealt{Arentoft2004}; \citealt{DeRidder2006}).

\section{Design}

Modularity principles lead to a design where each effect to be applied is separated in a different module so that it can be easily accessed and modified. Each of the noise effects applied to the image is generated in a different “Processing Step" module, containing the algorithm implementation in separated classes. The whole system is structured in separated components, being the “Processing Steps” one of them.  
	Modular design facilitates the processor architecture, separating the processing itself in three different components named “Preprocessing”, “Processing” and “Postprocessing”. Besides these components, there is a “Manager” module in charge of triggering, monitoring and controlling the whole system and a DataSet containing the images and processing parameters. Figure \ref{fig:architecture} shows an overview of the architectural design.

%%% Figure 2 %%%%%%%%%%%%%%%%%%%%%%%%%%%%%%%%%%%%%%%%%%%%%%%%%%%%%%%%%%%%%%%%%%%%%%%%%
\begin{figure*}[!ht]
\centering
\includegraphics[width=\textwidth, clip,angle=0]{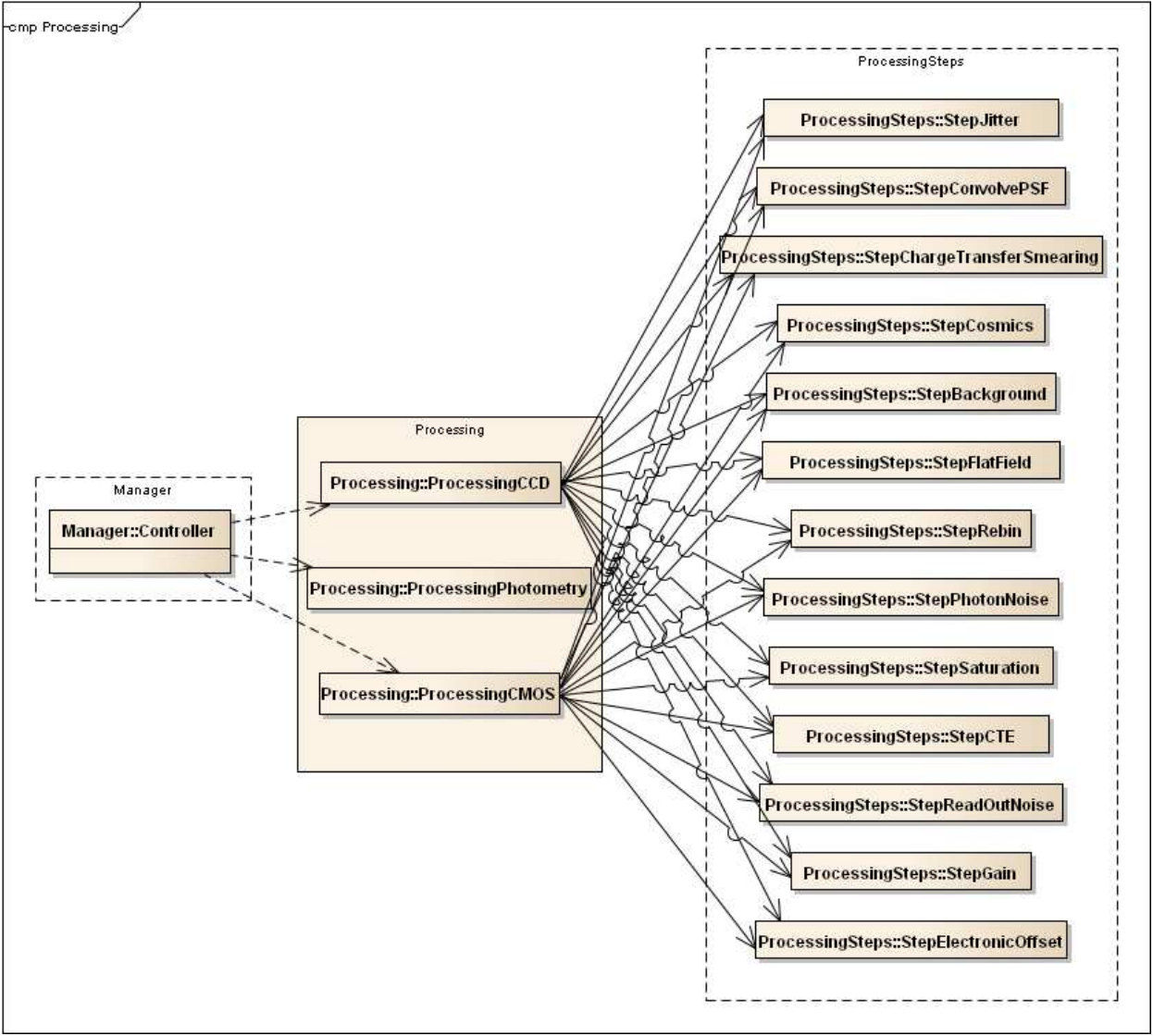}
\caption{Processing call diagram showing different simulation configurations using the same “Processing Steps” modules mimicking a pipeline (top to bottom).}
\label{fig:calldiagram}
\end{figure*}
%%% Figure 2 %%%%%%%%%%%%%%%%%%%%%%%%%%%%%%%%%%%%%%%%%%%%%%%%%%%%%%%%%%%%%%%%%%%%%%%%%

The type of processing to be performed is defined in the “Manager” component depending on the simulation selected by the user (CCD, CMOS, Photometric flux extraction,...). According to this selection, different classes are triggered in the “Preprocessing”, “Processing” and “Postprocessing” subsystems.
	The input parameters are defined in an XML-file and are structured according to their function in the simulator, e.g. Satellite, Telescope, Stellar field, CCD, or PSF. 
	The “Preprocessing” component is in charge of preparing the system for the processing. This implies that the “Preprocessing” subsystem reads all the input files and makes sure that all CCD/CMOS and any other parameters required by any of the processing steps (included in the “Processing Steps” component) are included in the DataSet. 
	As a part of the “Preprocessing” subsystem, we have included the “Parameters” component. This component performs all the calculations indicated by the “Preprocessing” subsystem to set the parameters required for the simulation into the DataSet. These parameters, related to the detector, the star field or any other parameter shared between exposures, are calculated in advance to the processing itself and loaded in the DataSet in order to avoid repeating blocks of code and save processing time. The parameters to be calculated in the “Parameters” component are determined by the Controller depending on the simulation to be performed. Given the CCD simulation example, the Parameters class to be used is the ParamsCCD. This class defines all the properties of the CCD, initializes the sub- and normal-pixel maps and contains methods and algorithms for dealing with the physical and electrical properties of the CCD.
	The DataSet module contains all the parameters required to perform the simulation and all the image maps. This includes from the input parameters read in the beginning of the simulation and the calculated parameters employed in the processing, to the final simulated images. This is intended to ease the access to each of the variables and intermediate product or image, in order to analyze every the processing step and its algorithms. 
	The “Processing Steps” component includes all the noise effects required in the simulation. Each of the effects is deployed in a different module, corresponding to a different step of the processing in such a way that different configurations may use each of the modules separately according to the needs. Figure \ref{fig:calldiagram} shows a call diagram including the “Processing” component and the classes implemented in it. As an use case example, the ProcessingCCD class calls every module in the "Processing Steps" component (and so does e.g. the ProcessingCMOS) as all the noise sources are intended to be applied in the baseline processing, but the ProcessingCCD class can easily be modified to leave out any of the noise sources. In the same way, new processing step modules can be implemented and added to any configuration process with a minor impact on the rest of the system. This modular design is a key factor in the design in order to aim for easy usability of this package for other missions.

The Photometry module (ProcessingPhotometry component in Figure \ref{fig:calldiagram}) is a different type of processing designed to analyze the image products generated in the CCD (or CMOS) processing. When the simulation process is configured to perform the photometry processing (through the Controller component as indicated in the input configuration), the ProcessingPhotometry is called in order to analyze the generated images by photometric algorithms and statistical tools to estimate the noise properties of the data. 

The main functionality of the “Postprocessing” component is to write to files the generated images and the required output parameters to disk. The required output is read from the DataSet module. Some external libraries might be required to write these datasets to disk.

\subsection{Imaging model}

To model the synthetic image of a science detector, one must take into account a set of parameters in order to ensure obtaining a realistic effect of the implicated noise sources. As a basic initial step, one must define the field of view (hereafter FoV) and take as input parameters the CCD size, number of pixels, pixel scale and project the position of the input star catalogue into this FoV. In order to accomplish computational requirements, only a part of the image is simulated and the obtained sub-field image results are extrapolated to the rest of the frame. 
	To increase the simulation accuracy, the calculations are applied at subpixel level. This means that each pixel is represented by a squared map of subpixels to take into account the intra-pixel sensitivity of the CCD to capture some of the effects such as the jittering (usually movements smaller than the pixel size) and the pixel sensitivity variations - flat field (assuming a 1/f spatial power spectrum which resembles a typical CCD as shown in \citealt{DeRidder2006}). Once the noise effects modeled at sub-pixel level are applied, the sub-pixels are rebinned back again to the original pixel scale. The processing pipeline corresponds to the "Processing Steps" right column top to bottom in Figure \ref{fig:calldiagram}.
	The PSF degradation is applied to the sub-field image (performing a convolution in the Fourier space) as well as the charge-transfer smearing, the cosmic hits and the sky background. Once the rebinning is applied to the image and is scaled back to the image, the photon noise, full-well saturation, charge-transfer efficiency, read-out noise, gain and the electronic offset effects are applied to the synthetic image as shown in Figure \ref{fig:calldiagram}.

%%% Figure 3 %%%%%%%%%%%%%%%%%%%%%%%%%%%%%%%%%%%%%%%%%%%%%%%%%%%%%%%%%%%%%%%%%%%%%%%%%
\begin{figure*}[!ht]
\centering
\includegraphics[width=0.9\textwidth,trim=10 0 10 30, clip,angle=0]{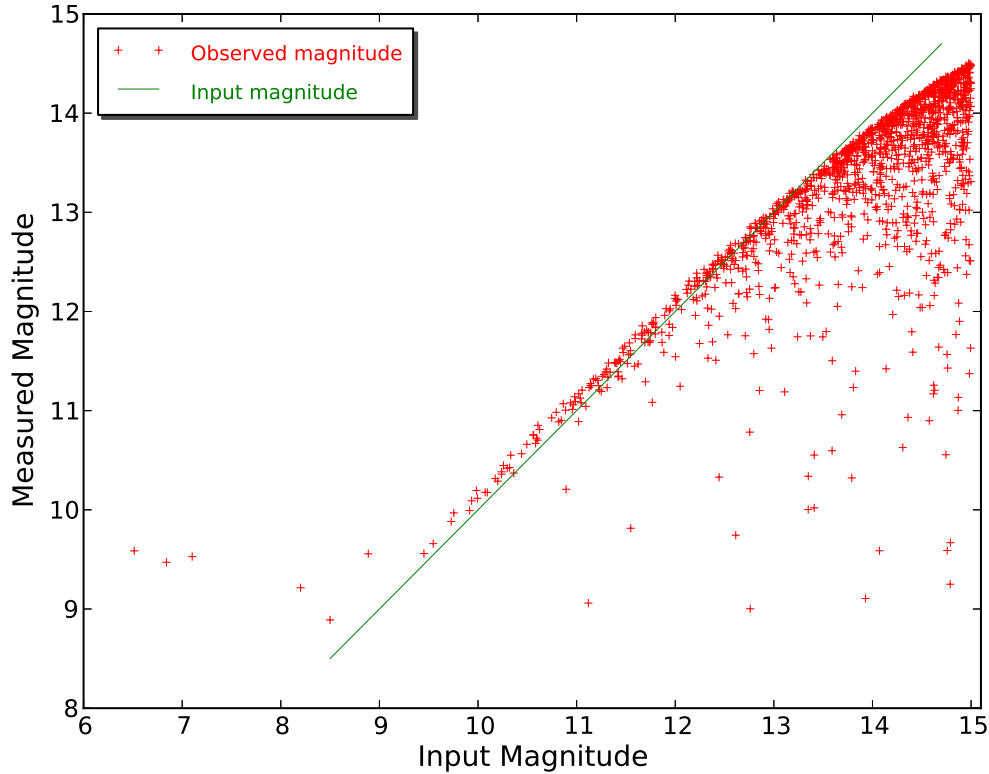}
\caption{Magnitude of the stars in the generated synthetic images as measured with the photometric process as a function of the input magnitude given in star catalogue. The green line indicates the measured magnitude equal to the input magnitude.}
\label{fig:graph}
\end{figure*}
%%% Figure 3 %%%%%%%%%%%%%%%%%%%%%%%%%%%%%%%%%%%%%%%%%%%%%%%%%%%%%%%%%%%%%%%%%%%%%%%%%

\subsection{Simulation example and conclusions}

Series of simulations have been made to test the performance of the photometric observations of the PLATO mission in some concrete aspects regarding the Jittering noise, PSF and CCD performance. For this task, we used a star catalogue of one of the proposed FoV of PLATO containing more than 32000 sources with mv ≤ 15 \citep{Barbieri2004}. As an example of one of the assessment simulations made in the performance test, we show the simulation of one-week of observations time-series, corresponding to 24,192 exposures, which have been computed and analyzed using the photometric algorithms for the assessment of the concrete input conditions of the mission. Figure \ref{fig:graph} presents an output of the analysis features of this simulation. It shows the obtained magnitude for each source detected with the photometry algorithms in the output synthetic images as a function of the magnitude of the same sources in the input star catalogue. The degradation in performance due to noise as a function of the magnitude is represented. In this plot, the sources with input mv ≤ 9 present measured magnitude brighter than the input magnitude since the flux of other stars leaks into the photometry mask.

In the future, we plan to expand the simulator modalities and perform various sets of simulations along with their analysis as illustration of the capabilities of our software package.

\acknowledgments
%________________________________________________________________
The research presented here was based on funding from the European Research Council under the European Community's Seventh Framework Programme (FP7/2007--2013)/ERC grant agreement n$^\circ$227224 (PROSPERITY) and from the Belgian federal science policy office Belspo (C2097-PRODEX PLATO Science Development).

\bibliographystyle{plainnat}
\bibliography{astroidsim}		% expects file "myrefs.bib"

\begin{thebibliography}{7}
\providecommand{\natexlab}[1]{#1}
\providecommand{\url}[1]{\texttt{#1}}
\expandafter\ifx\csname urlstyle\endcsname\relax
  \providecommand{\doi}[1]{doi: #1}\else
  \providecommand{\doi}{doi: \begingroup \urlstyle{rm}\Url}\fi

\bibitem[{Arentoft} et~al.(2004){Arentoft}, {Kjeldsen}, {De Ridder}, and
  {Stello}]{Arentoft2004}
T.~{Arentoft}, H.~{Kjeldsen}, J.~{De Ridder}, and D.~{Stello}.
\newblock {The Eddington CCD data simulator}.
\newblock In F.~{Favata}, S.~{Aigrain}, and A.~{Wilson}, editors, \emph{Stellar
  Structure and Habitable Planet Finding}, volume 538 of \emph{ESA Special
  Publication}, pages 59--64, January 2004.

\bibitem[Barbieri et~al.(2004)Barbieri, Piotto, Claudi, Crescenzio, Desidera,
  Baruffolo, Bedin, Bertelli, Gratton, Marzari, Montalto, and
  Ortolani]{Barbieri2004}
M~Barbieri, G~Piotto, R~U Claudi, G~Crescenzio, S~Desidera, A~Baruffolo, L~R
  Bedin, G~Bertelli, R~Gratton, F~Marzari, M~Montalto, and S~Ortolani.
\newblock {Search for an optimal Eddington Planet Finding Field}.
\newblock In F~Favata, S~Aigrain, and A~Wilson, editors, \emph{Second Eddington
  Workshop: Stellar structure and habitable planet finding}, pages 163 -- 175,
  Palermo, Italy, 9 - 11 April 2003, 2004. Noordwijk: ESA Publications
  Division.

\bibitem[Catala(2008)]{Catala2008}
Claude Catala.
\newblock {PLATO: PLAnetary Transits and Oscillations of stars}.
\newblock \emph{Experimental Astronomy}, 23\penalty0 (1):\penalty0 329--356,
  September 2008.

\bibitem[Claudi(2010)]{Claudi2010}
Riccardo Claudi.
\newblock {A new opportunity from space: PLATO mission}.
\newblock \emph{\apss}, 328\penalty0 (1-2):\penalty0 319--323, February 2010.

\bibitem[{De Ridder} et~al.(2006){De Ridder}, Arentoft, and
  Kjeldsen]{DeRidder2006}
J.~{De Ridder}, T.~Arentoft, and H.~Kjeldsen.
\newblock {Modelling space-based high-precision photometry for asteroseismic
  applications}.
\newblock \emph{\mnras}, 365\penalty0 (2):\penalty0 595--605, January 2006.

\bibitem[Rauer and Catala(2010)]{Rauer2010}
Heike Rauer and Claude Catala.
\newblock The plato mission.
\newblock \emph{Proceedings of the International Astronomical Union},
  6:\penalty0 354--358, 9 2010.

\bibitem[Zima et~al.(2010)Zima, Arentoft, {De Ridder}, Salmon, Catala,
  Kjeldsen, and Aerts]{Zima2010}
W~Zima, T~Arentoft, J.~{De Ridder}, S~Salmon, C~Catala, H~Kjeldsen, and
  C~Aerts.
\newblock {The PLATO End-to-End CCD Simulator -- Modelling space-based
  ultra-high precision CCD photometry for the assessment study of the PLATO
  Mission}.
\newblock 793\penalty0 (88):\penalty0 5, April 2010.

\end{thebibliography}

%\bibliographystyle{aa}	    % (uses file "plain.bst")
%\bibliography{platosim_aa}		% expects file "myrefs.bib"

\clearpage

%%% Use the figure environment and \plotone or \plottwo to include
%%% figures and captions in your electronic submission.
%%% To embed the sample graphics in
%%% the file, uncomment the \plotone, \plottwo, and
%

\end{document}